\documentclass[fleqn,usenatbib]{mnras}

\usepackage{newtxtext,newtxmath}
\usepackage{subfig}

\usepackage[T1]{fontenc}
\usepackage{ae,aecompl}
\usepackage[acronym]{glossaries}
\usepackage{hyperref}
\usepackage{comment}
\usepackage{newtxtext,newtxmath}
\usepackage{graphicx}	% Including figure files
\usepackage{amsmath}	% Advanced maths commands
\usepackage{epsfig}
\def\der{{\rm d}}
\DeclareUnicodeCharacter{2212}{-}

\title[21-cm signal using EDGES data]{Using Artificial Neural Networks to extract the 21-cm Global Signal from the EDGES data }

\author[M. Choudhury et al]{Madhurima Choudhury $^{1}$\thanks{E-mail: madhurimachoudhury811@gmail.com},
 Atrideb Chatterjee $^{2}$, 
Abhirup Datta $^{1}$,
\newauthor Tirthankar Roy Choudhury $^{2}$ 
\\
$^{1}$Discipline of Astronomy, Astrophysics Indian Institute of Technology Indore, India\\
$^{2}$National Centre for Radio Astrophysics, Tata Institute of Fundamental Research, Pune 411007, India\\
}

\date{Accepted XXX. Received YYY; in original form ZZZ}

\pubyear{2017}

\begin{document}
\label{firstpage}
\pagerange{\pageref{firstpage}--\pageref{lastpage}}
\maketitle

\begin{abstract}
The redshifted 21-cm signal of neutral Hydrogen is a promising probe into the period of evolution of our Universe when the first stars were formed (Cosmic Dawn), to the period where the entire Universe changed its state from being completely neutral to completely ionized (Reionization). The most striking feature of this line of neutral Hydrogen is that it can be observed across an entire frequency range as a sky-averaged continuous signature, or its fluctuations can be measured using an interferometer. However, the 21-cm signal is very faint and is dominated by a much brighter Galactic and extra-galactic foregrounds, making it an observational challenge. We have used different physical models to simulate various realizations of the 21-cm Global signals, including an excess radio background to match the amplitude of the EDGES 21-cm signal. First, we have used an artificial neural network (ANN) to extract the astrophysical parameters from these simulated datasets. Then, mock observations were generated by adding a physically motivated foreground model and an ANN was used to extract the astrophysical parameters from such data. The $R^2$ score of our predictions from the mock-observations is in the range of 0.65-0.89. We have used this ANN to predict the signal parameters giving the EDGES data as the input. We find that the reconstructed signal closely mimics the amplitude of the reported detection. The recovered parameters can be used to infer the physical state of the gas at high redshifts. 
\end{abstract}


%%%%%%%%%%%%%%%%%%%%%%%%%%%%%%%%%%%%%%%%%%%%%%%%%%

%%%%%%%%%%%%%%%%% BODY OF PAPER %%%%%%%%%%%%%%%%%%

\section{Introduction} 
The evolution of our Universe from the Cosmic Dawn (CD) and Epoch of Reionization (EoR) is not very well understood. The HI 21-cm line promises to be an excellent probe into these epochs \citep{furlanetto2006b, Morales_2010, Pritchard_2012}. The Global 21-cm experiments aim to measure the sky-averaged signature of this redshifted line, by using ground-based radio telescopes. Examples of such experiments are the Shaped Antenna measurement of the background RAdio Spectrum (SARAS), \citep{Patra_2013,Singh_2017}, the Large-Aperture Experiment to Detect the Dark Ages (LEDA) \citep{Greenhill_2012}, SCI-HI \citep{Voytek_2014}, the Broadband Instrument for Global Hydrogen Reionisation Signal (BIGHORNS) \citep{Sokolowski_2015}, 
and the Cosmic Twilight Polarimeter, CTP \citep{Nahn_2018}. The Experiment to Detect the Global EoR Signature (EDGES, \citet{Bowman_2018}) has reported a possible detection of the sky-averaged HI 21-cm Global signal from the Cosmic Dawn. However, this signal has an absorption trough of about $0.5$K, which is twice the amplitude predicted by the standard model of cosmology. If confirmed, this measured signal would give us a completely new insight into the physics of the evolution of the Universe. Following this detection, several models explaining this exceedingly deep absorption trough has been explored. \citep{Barkana_2018, fraser2018, pospelov2018, slatyer2018} had explained this with models with excess cooling from nonstandard physics, including dark matter particles scattering off baryons. \citep{Fialkov_2018, fraser2018, pospelov2018, EwallWice_2018, feng2018} have explored various possible 21-cm signals, varying the properties of the dark matter particles, in
addition to varying the astrophysical parameters.  \citet{EwallWice_2018,feng2018,fialkov2019}, have explored models which produce an excess radio
background, in addition to the cosmic radio background to explain the large amplitude of the absorption feature in the EDGES detection. This radio excess could be, for example, due to supernovae \citep{mirocha2019, jana2019} or primordial black holes \cite{EwallWice_2018} .

However, detecting the 21-cm signal is an observational challenge. When observed from Earth, the 21-cm signal, which is of the order of a few hundreds of $\mathrm{mK}$, is buried in a sea of very bright, dominating foregrounds. These foregrounds are several orders of magnitude brighter than the signal. In addition, there are effects of ionospheric distortion, radio frequency interference (RFI) and the frequency response of instrument which make the observed sky vary with time and other factors. One of the common techniques for extracting the faint cosmological signal is by assuming that the foreground is well-characterized, spectrally smooth and can be efficiently removed from the total signal. The residual left would contain the 21-cm signal including the signatures from the early phases of the formation of the Universe.\\

There have been several works on the application of machine learning (ML) techniques in the epoch of reionization. For example, \citet{Shimabukuro_2017} have used machine learning algorithms for extracting the parameters of the 21-cm power spectrum, while \citet{Schmit_2017} have built an emulator to generate several different realizations of 21-cm power spectrum using artificial neural networks (ANN). \citet{Cohen_2019} have emulated the 21-cm Global signal from the astrophysical parameters using ANN. \citet{Hassan_2019} have used convolutional neural networks (CNN) to identify reionization sources from 21-cm maps. Deep learning models have been used  to emulate the radiative transfer during the epoch of reionization in \citet{Chardin_2019}. In another work, \citet{Gillet_2019} have recovered the  astrophysical parameters directly from 21-cm images, using deep learning with CNN. \citet{Jennings_2019} have compared machine learning techniques for predicting 21-cm power spectrum from reionization simulations.
While all the above-mentioned works involve only the 21-cm signal of interest and reionization simulations, \citet{Choudhury_2020} had used artificial neural networks to extract the 21-cm Global signal parameters from mock observations which included effects of foregrounds, instrumental effects and noise. \\

In this paper, we use Artificial Neural Networks (ANNs), to extract the astrophysical parameters associated with the 21-cm Global signal from mock observational data. A simple code has been used to generate several 21-cm global signals by varying the input astrophysical parameters, which will be discussed in the following sections. Adding foregrounds to these 21-cm signals, we generate mock-observations which are then used as our training set. Using this, we design an ANN to predict the signal parameters from the total observed sky-signal with the added foreground. Finally, EDGES data has been used as an input to this ANN and we have obtained the predicted signal parameters. We then reconstruct the 21-cm signal using our code using the predictions.

This paper is structured as follows: We begin with a brief overview of the cosmological 21-cm signal in \textsection~2. In \textsection~3, we describe the physical models that we have used to simulate the 21-cm Global signal. The foreground model used in this work is described in detail in \textsection~4. In \textsection~5-6, we describe the methodology and some basics of artificial neural networks. We present our results obtained, in \textsection~7 and use the ANN to predict the 21-cm signal from actual data, in \textsection~8. Finally, we discuss the implications of our predictions and summarize our results.

\section{Computing the 21-cm Global Signal}
\label{sim_dTb}

The hyperfine transition line of atomic hydrogen arises from the hyperfine splitting of the ground state of the hydrogen due to the magnetic moment interaction of the proton and electron. The differential brightness temperature, $\delta T_{b}$, is measured relative to the Cosmic Microwave Background (CMB) temperature, $T_{\gamma}$ and is given by \citep{Furlanetto_2006}:
%\begin{equation}
%\delta T_{b} \equiv T_{b} - T_{\gamma}
%\end{equation}
\begin{equation}
\begin{split}
\delta T_{b}(\nu) & =\frac{T_{s}-T_{\gamma}}{1+z}(1-\exp^{-\tau_{\nu_{0}}})\\
& \approx 27x_{HI}(1+\delta_{b})\left(\frac{\Omega_{b}h^{2}}{0.023}\right)\left(\frac{0.15}{\Omega_{m}h^{2}} \frac{1+z}{10}\right)^{1/2}\\
& ~~\left(1-\frac{T_{\gamma}(z)}{T_{s}}\right)\Big[\frac{\partial_{r} v_{r}}{(1+z)H(z)}\Big]^{-1} ,
\end{split}
\label{eq:global}
\end{equation}
where $x_{HI}$ is the neutral fraction of hydrogen, $\delta_{b}$ is the fractional over-density of baryons, $\Omega_{b}$ and $\Omega_{M}$ are the baryon and total matter density respectively, in units of the critical density, $H(z)$ is the Hubble parameter and $T_{\gamma}(z)$ is the CMB temperature at redshift z, $T_{s}$ is the spin temperature of neutral hydrogen, and $\partial_{r} v_{r}$ is the velocity gradient along the line of sight. The observed 21-cm Global signal is nothing but the sky averaged differential brightness temperature, whose characteristic shape contains information about the physical state of the Universe at early times, e.g., the amount of UV radiation that can ionize hydrogen, the X-rays that can cause heating of the gas, and $Ly-\alpha$ which causes Wouthuysen-Field coupling \citep{Wouthuysen_1952,Field_1959}.  

In our calculations, we neglect the peculiar velocity term and the density fluctuation term
in the Global signal (Eqn.~\ref{eq:global}), as it averages out to a linear order and adds to a very small correction. So, Eqn.~\ref{eq:global} can be re-written as:
\begin{equation}
   { \delta T_{b}\approx 27(1-x_{i})\left(\frac{\Omega_{b}h^{2}}{0.023}\right) \left(\frac{0.15}{\Omega_{m}h^{2}} \frac{1+z}{10}\right)^{1/2} \left(1-\frac{T_{\gamma}}{T_{s}}\right)} 
\label{eq:dT_b}
\end{equation}
This equation is primarily used to construct the Global signal.

The spin temperature,  $T_{s}$,is the most interesting quantity in the expression for the differential brightness temperature which primarily determines the intensity of the 21-cm radiation [See Eqn.~\ref{eq:dT_b}]. The three competing processes that determine $T_{s}$ are: (1) absorption of CMB photons (as well as stimulated emission); (2) collisions with other hydrogen atoms, free electrons, and protons; and (3) scattering of Lyman alpha photons through excitation and de-excitation. The spin temperature is given by \citep{Field_1959, Pritchard_2015}:
\begin{equation}
T_{s}^{-1}=\frac{T_{\gamma}^{-1}+x_{c}T_{k}^{-1}+x_{\alpha}T_{\alpha}^{-1}}{1+x_{c}+x_{\alpha}},
\label{eq:SpinTemp}
\end{equation}
where $T_{\gamma}$ is the CMB temperature, $T_{k}$ is the kinetic gas temperature, $T_{\alpha}$ is the temperature related to the existence of ambient Lyman-alpha $(Ly\alpha)$ photons. $x_{c}$ and $x_{\alpha}$ are respectively the collisional coupling and the $Ly\alpha$ coupling terms. For a detailed review see \cite{Barkana_2005A}. \\
For this work, we developed a semi-numerical code which can produce a Global 21-cm signal over the redshift range $6<z<50$ in less than a second \citep{Chatterjee_2019}.  Our calculation closely follows that of \citet{furlanetto2006c} and \citet{pritchard2012}. It takes the following astrophysical parameters as input:
\begin{enumerate}
\item star formation efficiency $f_{\star}$,
\item the escape fraction $f_{esc}$ of ionizing photons,
\item X-ray heating efficiency $f_{X,h}$,
\item  the number of Ly$\alpha$ photons $N_{\alpha}$ produced per baryon in the relevant frequency range,
\item efficiency parameter $f_{R}$ of the radio background (only for models with excess radio background),
\end{enumerate}
and provides $\delta T_b$ as a function of redshift (or observed frequency). We discuss the parameters in more detail below.

% ************************************************
\subsection{The coupling coefficients}
% ************************************************

We will first discuss how the coupling coefficients are calculated.

\begin{enumerate}

\item There are three different channels of collisions, e.g., Hydrogen-Hydrogen (H-H), Hydrogen-electron (H-e) and Hydrogen-proton (H-p), which determine the collisional coupling coefficient, $x_c$. The calculation of the coupling coefficient in our numerical code is described in detail in \cite{Chatterjee_2019}.

\item The  Ly$\alpha$ coupling coefficient, $x_{\alpha}$ can be computed once the background Ly$\alpha$ flux, $J_{\alpha}$, is known \citep{furlanetto2006b} (F06 hereafter). We approximate $J_{\alpha}$ following F06.
\end{enumerate}

% ************************************************
\subsection{The gas kinetic temperature}
% ************************************************

In the redshift regime of our interest, the gas kinetic temperature, $T_{k}$, is governed by two processes {\it (i)} Adiabatic cooling due to expansion of the Universe and {\it (ii)} X-ray heating (F06). Given our lack of understanding in modelling the X-ray background we take the simplest possible approach as described in F06. We simply extrapolate the correlation between the star formation rate (SFR), $\dot{M}_*$ and the X-ray luminosity, $L_X$, in the present Universe to higher redshifts and write it as,
\begin{equation}
    L_X=3.4 \times 10^{33} f_X \left(\frac{\dot{M}_*}{\text{M}_{\odot}~\text{yr}^{-1}} \right) \mbox{J~s}^{-1},
\end{equation}
where $f_X$ is an unknown normalization factor taking into account the difference between local and high-$z$ observations. Defining $f_{coll}$ to be the collapsed fraction of dark matter halos with mass greater than $m_{min}$, the X-ray emissivity can be expressed as,
\begin{equation}
 \epsilon_X=7.5 \times 10^3 \text{K} \, n k_B H(z)\times f_{X,h}  \left(\frac{f_{\star}}{0.1}  \frac{d f_{coll}/d z}{0.01} \frac{1+z}{10} \right) ,
\end{equation}
where n is the total number of gas particles, $k_B$ is the Boltzmann’s constant and $f_{X,h} = f_X \times f_h$, where $f_h$ is the heating fraction of the X-rays (rest of them ionize the IGM) . 

Although it is possible to use a fixed value for the ionization fraction $x_i \simeq 10^{-4}$ \citep{bharadwaj2004} for calculating the 21-cm signal in the redshift range we are interested in, we still compute the evolution of $x_i$ with redshift using the following equation (F06), 
\begin{equation}
\frac{\der x_{i}}{\der t} = \zeta(z) \frac{d f_{coll}}{dt} - x_{i}~ \alpha_{B}~{\cal C}~n_{H,{\rm com}}~(1+z)^{3},
\end{equation}
where $n_{H, {\rm com}}$ is the comoving number density of Hydrogen, $\alpha_B$ is the (case B) recombination rate coefficient and ${\cal C}$ is the clumping factor of the IGM. $\zeta(z)$ is defined as,
\begin{equation}
\zeta=A_{\mathrm He}f_{\star}f_{esc}\text{N}_{\text{ion}}
\end{equation}
where $A_{\mathrm He}=1.21$ is a correction factor for helium, $f_{\rm esc}$ is the escape fraction of ionizing photons, $\text{N}_{\text{ion}}$ is the number of ionizing photons per baryon produced in stars. We modelled the clumping factor as ${\cal C}= 1 + 43z^{−1.71}$ \citep{haardt2012,pawlik2009}.

\begin{figure*}
\subfloat[Traditional set]{\includegraphics[width=0.33\textwidth]{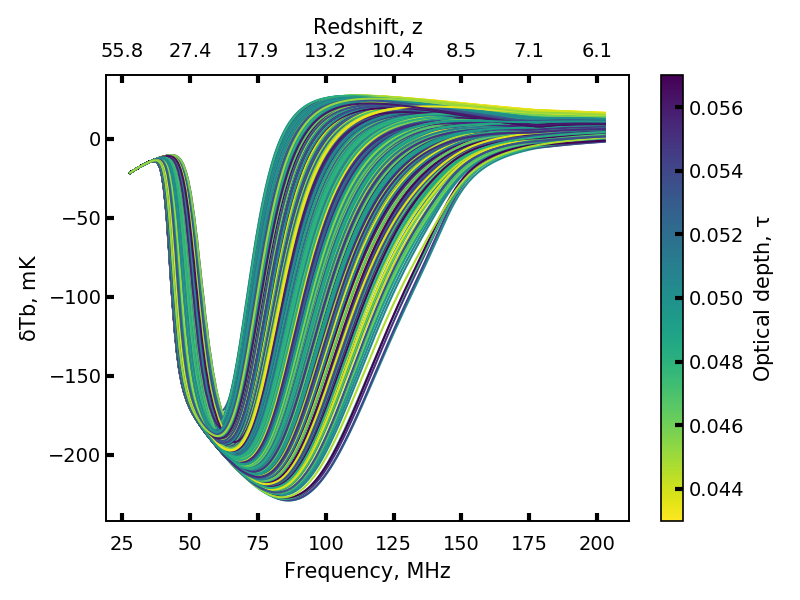}}
\subfloat[Exotic Set]{\includegraphics[width=0.33\textwidth]{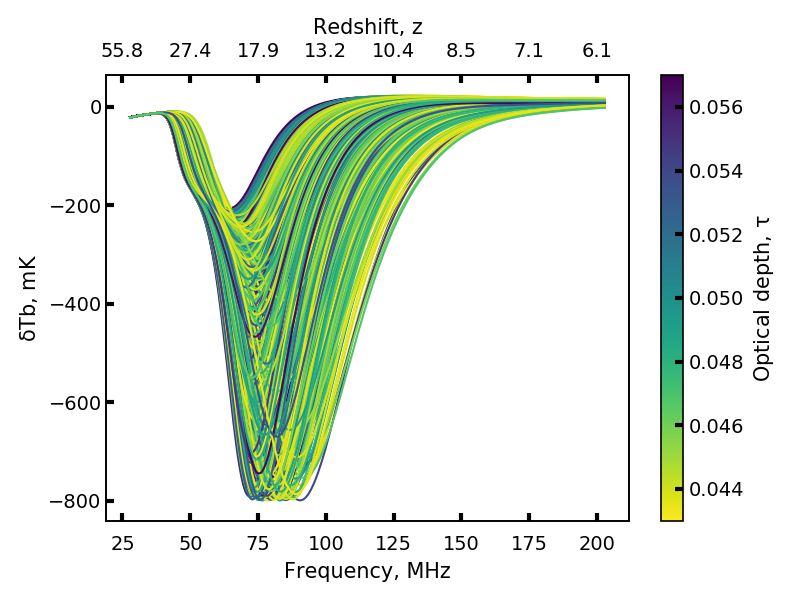}}
\subfloat[Total set]{\includegraphics[width=0.33\textwidth]{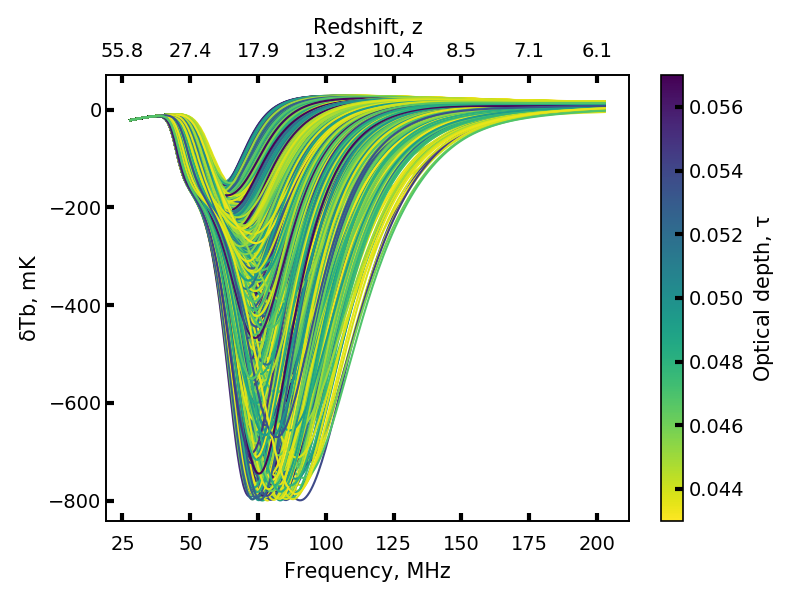}} 
\caption{The sample 21-cm signals generated by using the (a)traditional models and (b)the models with excess radio background. The differential brightness temperature, $\delta T_b$, is plotted versus frequencies(MHz). This set of sample 21cm signals are used for building the training datasets (c). The colorbar on the right represents the optical depth.}
\label{fig:21-cmsignal_sets}
\end{figure*}

\subsection{Models with excess radio background}
\label{excess}
% ************************
It has now been well accepted that matching the amplitude of the EDGES signal would either require making the gas colder, using exotic physics such as new dark matter interactions \citep{barkana2018, fraser2018, pospelov2018, slatyer2018} or invoking a radiation temperature larger than the CMB temperature \citep{fraser2018,pospelov2018,ewall_wice2018,feng2018}. In this work, we take the latter approach as it got some support by observations of an excess radio background in the local Universe from ARCADE-2 experiment \citep{fixsen2011}. 

The excess radio background is modelled assuming the local radio-SFR ($L_{R}-\dot{M}_*$) relation at $150$ MHz \citep{gurkan2018} holds at higher redshift and can be written as 
\begin{equation}
L_R = 10^{22} f_R \left(\frac{\dot{M}_*}{1\text{M}_{\odot}\text{yr}^{-1}}\right)\mbox{J~s}^{-1}\mbox{Hz}^{-1} ,
\end{equation}
where $f_R$ is a free parameter taking into account any differences between the local observations and the observations of the high redshift Universe. This relation is extrapolated to higher frequencies by assuming a spectral index of $-0.7$ \citep{gurkan2018}. The globally averaged radio luminosity at 150 MHz per unit comoving volume at redshift $z$ is then given by,
\begin{equation}
  \epsilon_{R,150}(z) = f_R \times 10^{22} f_{\star} {\bar{\rho}}_{\text{m}} \frac{\der f_{coll}}{\der t} \, {\rm J~ s^{-1} Hz^{-1} Mpc^{-3}},
  \end{equation} 
therefore the corresponding 21-cm radiation flux can be computed as \citep{2003ApJ...596....1C}
\begin{equation}
\begin{aligned}
F_{R}(z)= &  \left(\frac{1420}{150}\right)^{-0.7} \frac{c(1+z)^{3.7}}{4\pi} f_R \\
& \times 10^{22} f_{\star} {\bar{\rho}}_{\text{m}} \int_z^{\infty}  \frac{\der f_{coll}}{\der z'} \frac{1}{(1+z')^{0.7}} \der z',
\end{aligned}
\end{equation}
where ${\bar{\rho}}_{\text{m}}$ is the mean density of matter in the Universe. We convert this flux into a radio brightness temperature $T_{R}$. This results in a total background temperature given by $T_{\gamma}(z)=T_{R}(z)+T_{\rm CMB}(z)$. 

Although an excess radio background can enhance the amplitude of the absorption signal, a redshift independent value creates an radio excess which is much higher than that observed by ARCADE-2 experiment, hence this excess radio background has been turned off at $z = 16$ (however for numerical stability we model this as a tanh function) as discussed in \cite{Chatterjee_2019}. Physically, the rise and suppression of the radio background in a small redshift range could be explained by a transient population of metal-free PopIII stars \citep{2020MNRAS.493.1217M} or a rapid heating by cosmic rays \citet{jana2019}.

\begin{figure}
\includegraphics[width=\columnwidth]{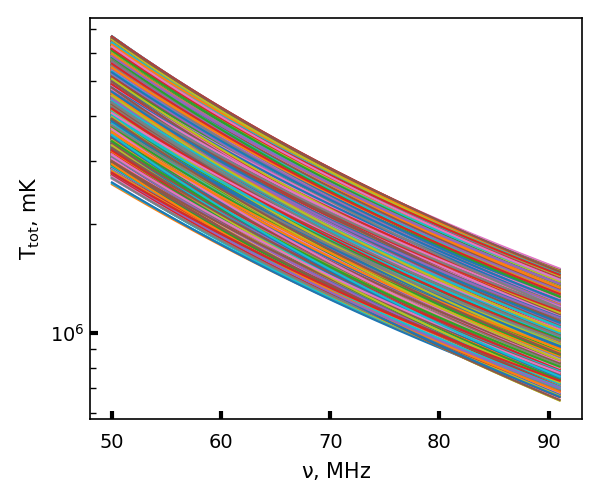}
\caption{The training dataset for the neural network. Here, we have plotted the total signal (21-cm signal along with foregrounds) which is used for training. As expected, the 21-cm Global signature is lost entirely when the foregrounds are added. The y-axis represents the total sky-signal in mK and on the x-axis, the corresponding frequency channels are shown. }
\label{fig:trainingset}
\end{figure}

\section{Foregrounds}
\label{foregrounds}
The major observational challenge while detecting the faint global 21-cm signal from CD/EoR is the bright foregrounds in the sky. Radio emission from our galaxy and other extra-galactic sources are several orders of magnitude brighter than the cosmological signal of interest. The expected global 21-cm signal is about four orders of magnitude weaker than the foreground emission. For all ground-based observations, there is a significant amount of corruption due to the variations in the Earth's atmospheric conditions as well. In addition, the instrument response further corrupts the signal. The bandwidth corresponding to the observations of the redshifted 21-cm signal overlaps with the bandwidth used for telecommunication, including the FM-radio band. All such man-made terrestrial contributions constitute the Radio Frequency Interference (RFI). The bright foregrounds, RFI, instrumental calibration errors, combined together  poise significant challenges to the Global 21-cm experiments. Hence, several sophisticated simulations and modelling are necessary to understand the effect of these corruption terms on the possible signal extraction methods. It also makes it critical to have an accurate model for the foregrounds at these radio frequencies, and a well calibrated instrument. A simple representation of the foregrounds as a polynomial in $\mathrm{\log(\nu)-\log T}$, is given in \citet{Pritchard_2010, Bernardi_2015}. In our work, we use the foreground model as in \citet{bowman2018}, which we call the Bowman-Rogers (BR) foreground model, Which we explained briefly in the following section. 

\subsection{Bowman-Rogers(BR) foreground model}
The observed sky brightness can be written as (following \citet{Hills_2018}):
\begin{equation}
    \rm{T_{sky}=(T_{BG}+T_{21}+T_{FG})e^{-\tau_{ion}}+T_e(1-e^{-\tau_{ion}} }),
\end{equation}
where, $\rm T_{BG}$ represents the background radiation which is arising from higher redshifts, $\rm T_{21}$ is the 21-cm signal feature, $\rm T_{FG}$ is the foreground radiation, $\rm \tau_{ion}$ is the opacity of the ionosphere and $\rm T_e$ is the opacity weighted temperature of the electrons in the ionosphere.
\citet{Bernardi_2015} use a log-polynomial expansion which describes the foreground emission as a function of frequency, around a central frequency or reference frequency $\rm \nu_c$, 
\begin{equation}
    \log \mathrm{T_{FG}(\nu)= \sum^{N-1}_{n=0}c_n[\log(\nu/\nu_c)]^n .}
\end{equation}

A linearized version of the physically-motivated foreground model, as used in \citet{Bowman_2018, Hills_2018}, along with the ionospheric terms, which we call the Bowman-Rogers (BR) foreground model $\rm{T_{FG(BR)}}$, is given by:
\begin{equation}
\begin{split}
\mathrm{T_{FG(BR)}} =~ & b_0 \cdot \left(\frac{\nu}{\nu_c}\right)^{-2.5} + b_1 \cdot \left(\frac{\nu}{\nu_c}\right)^{-2.5} \ln\left(\frac{\nu}{\nu_c}\right) \\
& + b_2 \cdot \left(\frac{\nu}{\nu_c}\right)^{-2.5}\cdot \left[\ln \left(\frac{\nu}{\nu_c}\right)\right]^2  \\
& + b_3 \cdot \left(\frac{\nu}{\nu_c}\right)^{-4.5}+ b_4 \cdot \left(\frac{\nu}{\nu_c}\right)^{-2} ,
\end{split}
\label{eq:fgBR}    
\end{equation}

where, the coefficients $\rm b_0,b_1,b_2,b_3,b_4$ are the parameters of the foreground model and $\rm \nu_c$ is a reference frequency which is taken to be 80 MHz for this work. We have used $\rm b_i$'s as the variables for the coefficients, instead of $\rm a_i$'s as used in \citet{Hills_2018,Bowman_2018}. We use this model for generating the foregrounds and construct the mock observations. We fit this model to a sky-averaged total spectra and find that the fitted values corresponding to a sub-band $\rm \approx(52.7-98.4)$ MHz are, $\rm \{b_0,~b_1,~b_2,~b_3,~b_4\} = \{-13491.45, -16643.00, -11561.28, 147.30, 14826.74\}$. 
Using these best-fit values, we vary the parameters of this foreground model by $\rm 1-5\%$ to generate the mock foreground data which is used to build the training sets(described in \textsection~\ref{TrainingSets}-\ref{Building_test_sets}. \\    
\begin{figure}
    \includegraphics[width=\columnwidth]{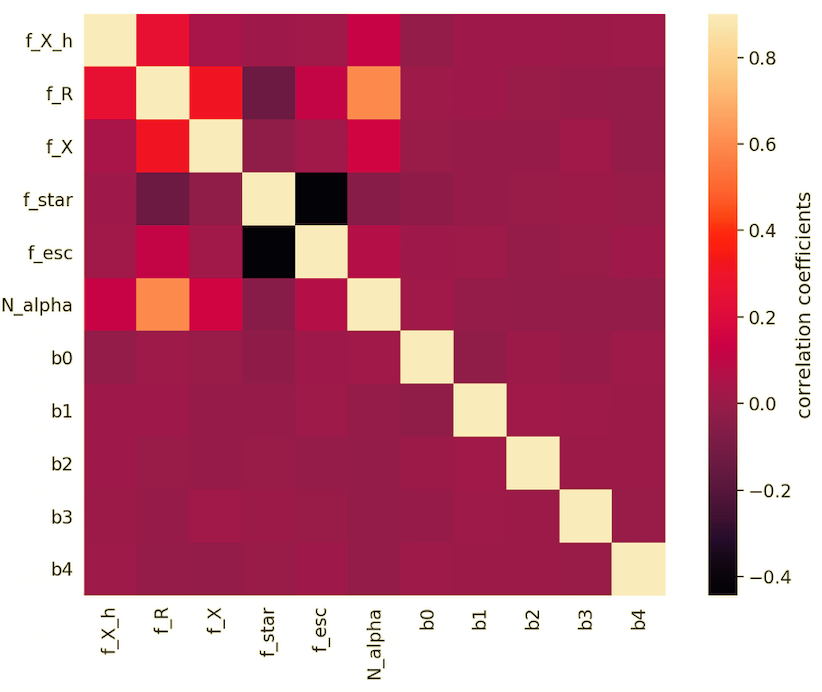}
    \caption{Correlation map of the various parameters used for training. We can observe from the plot that the astrophysical parameters, for example $\rm f_{esc}$ and $\rm f_*$ are highly correlated. The foreground parameters of this model are comparatively less correlated with the signal parameters.   }
    \label{fig:corr_map}
\end{figure}

\section{Overview of ANN}
In this section, we briefly introduce the concepts of ANN. The functional unit of an ANN is called neuron. A simple neural network consists of three basic layers: an input layer, one or more hidden layers, and an output layer. In a feed-forward, fully connected neural network, each neuron in a layer is connected to every neuron in the next layer and the flow of information is unidirectional. Each of these connections are associated with a weight and a bias. At the output layer, after each forward-pass a cost function or an error function is computed. This cost function is optimized, during the training process by back-propagating the errors, iteratively. A detailed description of the basic algorithm used in artificial neural networks is discussed in \citet{Choudhury_2020}. 
In our feed forward network, we have used an ANN with 4 Dense layers (using the Sequential Model from the \textsc{keras api}, Python). We have 1024 neurons in the input layer, corresponding to the 1024 frequency channels between $\rm 50-91~\mathrm{MHz}$. We have used standard \textsc{scikitlearn} \citep{Pedregosa_2011} and \textsc{keras} modules to build our network. We choose the number and size of the hidden layers for the best network performance. The number of neurons in the output layer corresponds to the number of output parameters which we want to predict. The detailed structure of the neural network used in our work is described in the following sections. 
\begin{table}
    \centering
    \begin{tabular}{c c c }
    \hline
     Parameters & Ranges  \\
     \hline
     $\rm f_X$ & 0.51-20.0 & \\
     $\rm f_{Xh}$ & 0.050-0.399 & \\
     $\rm f_R$ & 0  &(traditional) \\
      & 2000-18000 &(exotic) \\
     $\rm f_*$ & 0.0030-0.0099 &  \\
     $\rm f_{esc}$ & 0.06-0.19  & \\
     $\rm N_{\alpha}$ & 9000-80000 &  \\
    \hline
    \end{tabular}
\caption{Parameter range randomly sampled to generate the  21-cm signals for the training set using our code.}  
\label{tab:param_range}
\end{table}

\section{Building the training datasets}
\label{TrainingSets}
Using the models described in \textsection\ref{sim_dTb}, we generate several 21-cm Global signals by varying the signal parameters: $~\mathrm{\{f_X,f_{X,h},f_R,f_*,f_{esc},N_{\alpha}\}}$. The ranges of parameters used are tabulated in Tab.~\ref{tab:param_range}. We explicitly construct two sets of 21-cm Global signals: one set including only the traditional models, limiting the electron optical depth, $\rm \tau\approx 0.05$ and allowing a maximum absorption trough of $\rm -250\mathrm{mK}$ [Fig.\ref{fig:21-cmsignal_sets}]. The other set includes the models which account for the X-ray heating and an additional excess radio background. The amplitude of the absorption trough is limited to a maximum of $\rm -800 \mathrm{mK}$ with $\rm \tau\approx 0.05$. We take into account the frequency range $\rm 50-91$MHz, across 1024 channels to  build the total 21-cm signals set, which is then used to construct the training data. We construct this set in this manner to ensure that both the varieties of shapes of signals contribute equally to the training process, and we do not get a biased result. We call this set of signals the Global signal training set, and use an artificial neural network to predict the astrophysical parameters, without any added foregrounds in \textsection~\ref{Nofg}\\

The mock-data training set, consists of $\mathrm{T_{21}}$ along with the BR foreground, $\mathrm{T_{FG(BR)}}$ as given by Eqn.~\ref{eq:fgBR}. Each sample in the mock-data training set (see Fig.~\ref{fig:trainingset}) is given by:
\begin{equation}
    \mathrm{T_{tot}(\nu)=T_{21}(\nu)+T_{FG(BR)}(\nu)}
\label{eq:mockdatatrain}
\end{equation}
where, all the temperatures are in mK. 
The output parameter set consists of 6 signal parameters, $\mathrm{{f_X,f_{X,h},f_R, f_*,f_{esc},N_{\alpha}}}$ and 5 foreground parameters: $\mathrm{b_0,b_1,b_2,b_3,b_4}$. Fig.~\ref{fig:corr_map} shows the correlation between the various parameters, which we are using for training the network. We see that the astrophysical parameters, as expected, are strongly correlated as compared to the foreground parameters. The ranges of the parameters explored are listed in the Tab.~\ref{tab:param_range}.

\section{Building the test sets}
\label{Building_test_sets}
For this work, we construct two test sets to check the performance of our network. 
(i) Test Set A: Consists of 1000 sample 21-cm signals from both categories, along with the BR foregrounds and noise corresponding to 1000 hours of observation. (ii) Test Set B:  Includes one sample 21-cm signal which contains the excess radio background and one traditional 21-cm signal (without any excess radio background). Both the samples are contaminated by the BR foregrounds and noise, corresponding to 1000 hours of observation.
The test sets are constructed as:
\begin{equation}
    \mathrm{T_{test} = T_{21} + T_{FG(BR)} + \sigma_{t}}
\end{equation}
Here, $\rm \sigma_t$ is the thermal noise corresponding to a bandwidth, $\rm \Delta \nu$ and an observation time, $\rm \tau $. From the standard radiometer equation, $\rm \sigma_{t}$ can be re-written for $\rm N_t$ hours of observation:
\begin{eqnarray*}
    \sigma_t(\nu) &=&\frac{T_{sys}(\nu)}{\sqrt{\Delta\nu\cdot\tau}}\\
    &=&\frac{T_{FG}(\nu)}{\sqrt{\Delta\nu\cdot 10^6\cdot 3600\cdot N_t}}
\end{eqnarray*}
After the network has been saved, we use these test sets to predict the output parameters (6 signal and 5 foreground parameters). 
The test set A consisting of 1000 sample 21-cm signals, is used to compute the performance metric of the network, while test set B is used to check the accuracy of the signal parameter extraction. 

\section{$\mathrm {R^2}$ Score}
We choose $R^2$-score as a metric to assess and compare the performance of the networks. The coefficient of determination, $\rm R^2$, is calculated for each of the parameters from the predictions of test set A. The $\rm R^2$ score is defined as:
\begin{equation}
\mathrm{R^2=\frac{\Sigma(y_{pred}-\overline{y}_{ori})^2}{\Sigma(y_{ori}-\overline{y}_{ori})^2}=1-\frac{\Sigma(y_{pred}-y_{ori})^2}{\Sigma(y_{ori}-\overline{y}_{ori})^2}}
\end{equation}
where, $\overline{y}_{\mathrm{ori}}$ is the average of the original or true parameter, and the summation is over the entire test set. The score $\rm R^2 =1$, implies a perfect inference of the parameters, while $\rm R^2$ can vary between 0 and 1. 

\section{Results}
\subsection{Training without foregrounds}
\label{Nofg}
\begin{figure*}
\centering
\subfloat[$\rm f_{R}/10^3$]{\includegraphics[width=0.33\textwidth]{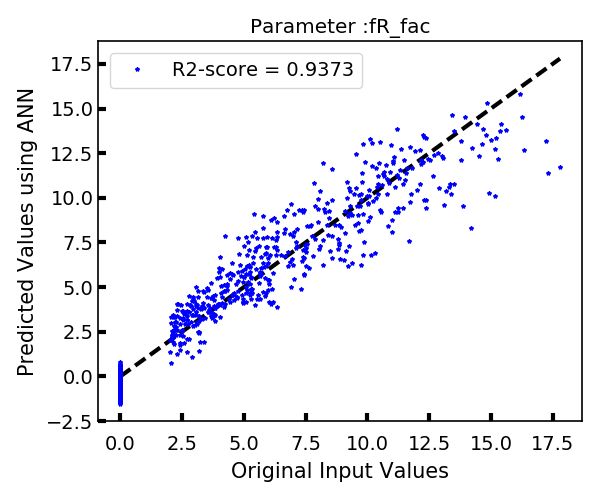}}
\subfloat[$\rm log~f_{*}$]{\includegraphics[width=0.33\textwidth]{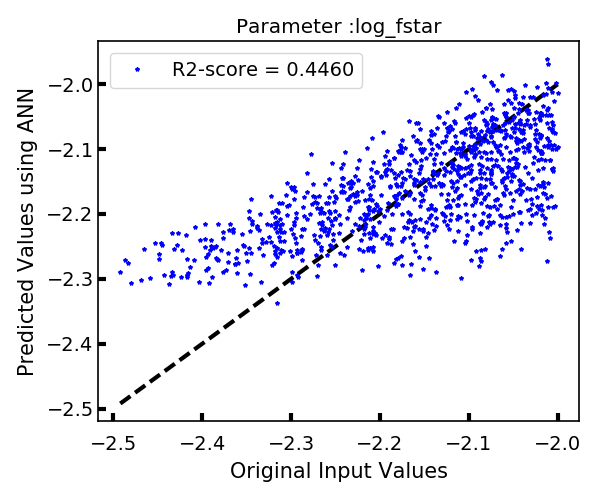}}
\subfloat[$\rm f_{esc}$]{\includegraphics[width=0.33\textwidth]{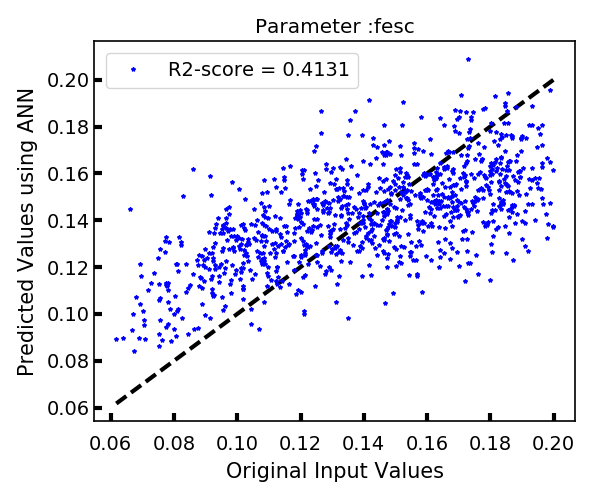}}
\\
\subfloat[$\rm f_{X}$]{\includegraphics[width=0.33\textwidth]{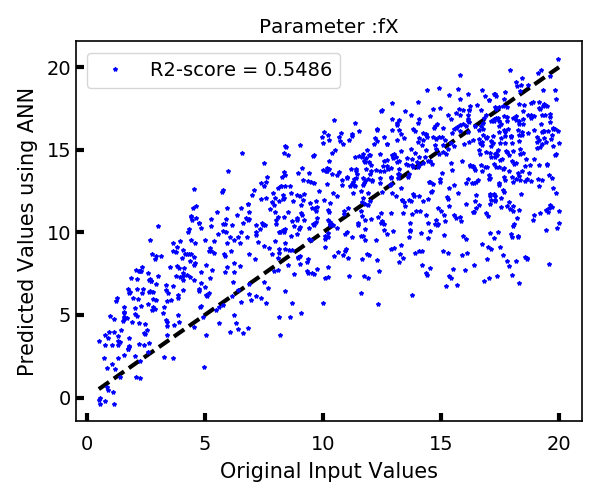}}
\subfloat[$\rm f_{Xh}$]{\includegraphics[width=0.33\textwidth]{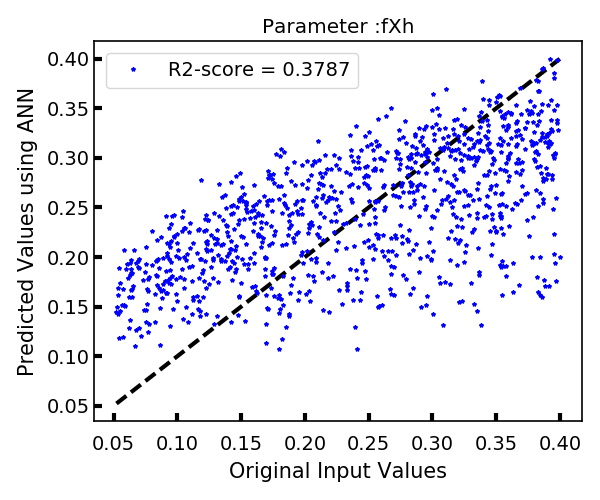}}
\subfloat[$\rm log~N_{\alpha}$]{\includegraphics[width=0.33\textwidth]{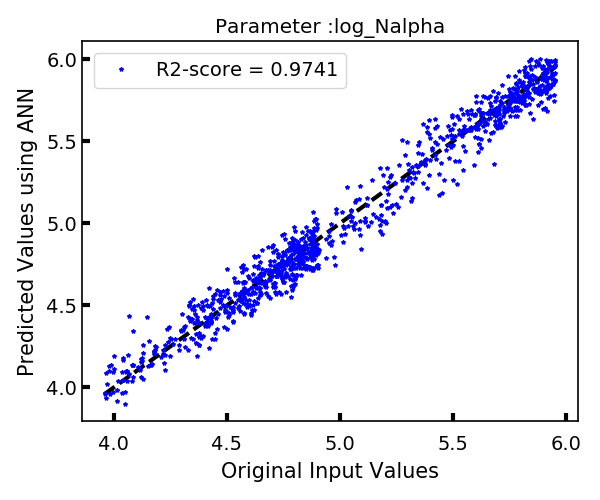}}
\caption{Each of the plots above is a scatter plot of the predicted values of the signal parameters, where the test test set consists of only 21-cm Global signals and noise for the case where the network is trained by only 21cm Global signals. The central dashed line through the origin depicts the true values. The R2 scores for each parameter is calculated and mentioned in the top corner in each plot.}
\label{fig:NoFg}
\end{figure*}

Firstly, we consider only the Global signal training set (as shown in Fig.~\ref{fig:21-cmsignal_sets}(c). We use the \textsc{Sequential API} from keras to build an ANN, with 1024 input neurons, corresponding to the 1024 frequency channels,  two hidden layers of 512 and 32 neurons respectively, activated by the 'sigmoid' activation function. the output layer has 6 output neurons, to predict the astrophysical parameters. The input is preprocessed by using the 'StandardScaler' function available in sklearn, while the signal parameters are scaled by the 'MinMaxScaler'. We use 'adagrad' \citep{Duchi_2011} as the optimizer and 'mean squared error' as  the loss function.  Once the network is trained, validated and tested, we save the model.
To constitute the test set for this case, we add thermal noise, $\rm \sigma_t$ corresponding to 1000 hours of observation, Which can be written as: 
\begin{equation}
    \rm{T_{test}'=T_{21} + \sigma _t} .
\end{equation}
We calculate the R2-scores for each parameter from the test set predictions. The predicted values of the parameters are shown in Fig.~\ref{fig:NoFg}. From the R2- scores we can observe that the parameters $\rm N_{\alpha}$ and $\rm f_R$ have the highest R2 scores of 0.9666 and 0.9239, respectively. However, $\rm N_{\alpha}$ is plotted in logarithmic scale, and $f_R$ is scaled by a factor of 1000. These two parameters are not as highly correlated as the $\rm f_{X}-f_{Xh}$ and $\rm f_*-f_{\rm esc}$ pairs. 

\begin{figure*}
\centering
\subfloat[$\rm f_{R}/10^3$]{\includegraphics[width=\columnwidth]{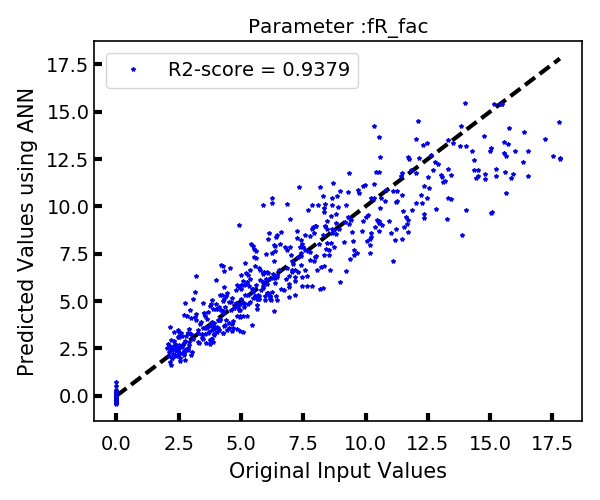}}
\subfloat[$\rm f_{*} \cdot f_{esc}$]{\includegraphics[width=\columnwidth]{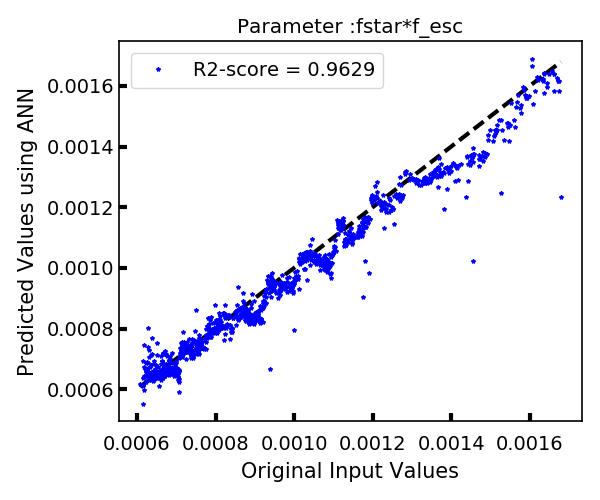} }
\\
\subfloat[$\rm f_{X} \cdot f_{Xh}$]{\includegraphics[width=\columnwidth]{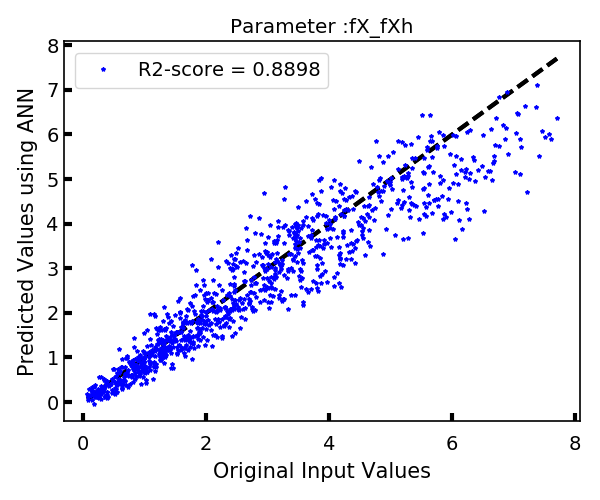}}
\subfloat[$\rm N_{\alpha}/10^3$]{\includegraphics[width=\columnwidth]{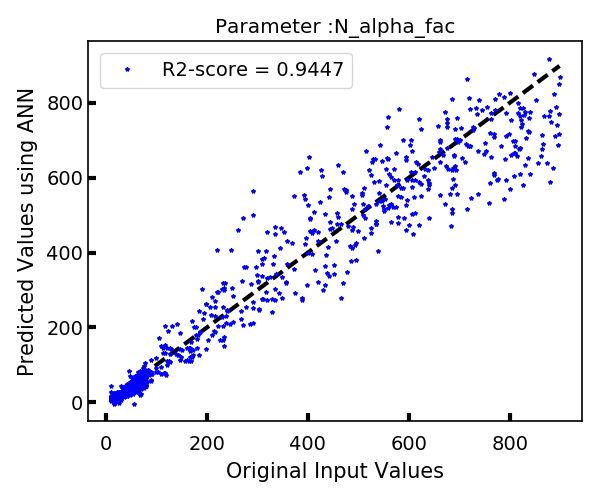}}
\caption{The plots above show the predicted values of the parameters for the test set consisting of only the 21-cm Global signal and noise. The network is trained only with 21cm Global signals (without adding foregrouds). We have combined the parameter-pairs which are highly correlated into single parameters. The first plot on the top panel shows the parameter $\rm f_R/1000$, and the second one is the combined parameter $\rm f_*\cdot f_{esc}$. The bottom panel shows the plots of the parameters $\rm f_{X} \cdot f_{Xh}$ and $\rm N_{\alpha}/1000$. The dashed straight line in each plot represents the true values of the parameters. The calculated R2-scores are mentioned in each plot.}
\label{fig:NoFg_combined}
\end{figure*}

In order to check whether the network performance improves, we combined the highly correlated parameters into a single parameter. So now the four output parameters would be $\rm f_R, f_* \cdot f_{esc}, f_x \cdot f_{xh}, N_{\alpha}$. We use a neural network with 1024 input neurons, two hidden layers with 256 and 133 neurons, respectively. The hidden layers neurons are activated by a 'tanh' function, and we use the 'adagrad' optimizer. The output layer consists of 4 neurons. We use the same test set as described above, and plot the predictions in Fig~\ref{fig:NoFg_combined}.  The   R2-scores obtained  are around $0.88-96$, for the combined parameters. These are much higher as compared to the previous case, where we had considered all the 6 signal parameters separately. This is expected, as we are well aware of the high correlation between these parameters.

\begin{figure}
    \centering
    \includegraphics[width=\columnwidth]{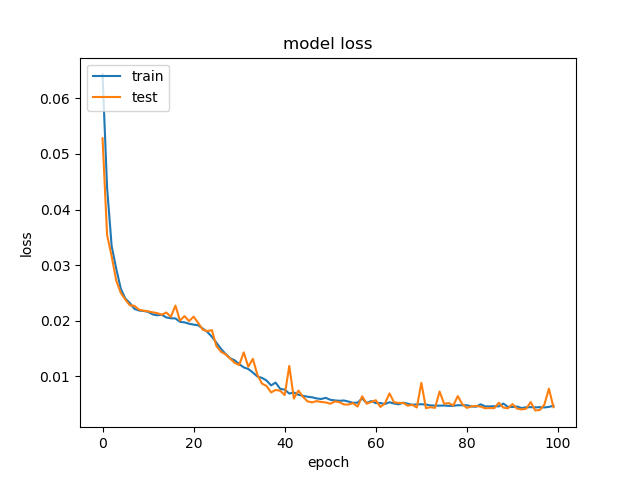}
    \caption{Evolution of the loss function for the network when there are no added foregrounds. The MSE is calculated at the end of every epoch, and optimized. The blue curve is the loss function for the traiing set which saturates around 80 epochs. The Loss function for the test set, closely follows that of the training set.  }
    \label{fig:my_label}
\end{figure}

\subsection{Training after adding Foregrounds}
Observations of the 21-cm signal would be totally dominated by foregrounds for all ground-based experiments. In this section, we have considered the training data to be the mock-data, as described by Eqn.~\ref{eq:mockdatatrain}. We use the \textsc{sequential} API form keras, to build the neural network. For the training dataset (as shown in Fig.~\ref{fig:trainingset}), we build a 4 layer neural network. The input layer has 1024 neurons corresponding to the number of the input frequency channels. There are 2 hidden layers with 256 and 64 neurons respectively. We use the sigmoid activation function in the hidden layers. The output layer consists of 11 neurons corresponding to the 6 signal parameters and the 5 foreground parameters. We use the `adam' \citep{Kingma_2014} optimizer, and mean squared error as the loss function. Once the network is trained and the model is saved, we use it to predict the parameters of the signal from the constructed test datasets. 

\begin{figure*}
\centering
\subfloat[$\rm f_{X}$]{\includegraphics[width=0.33\textwidth]{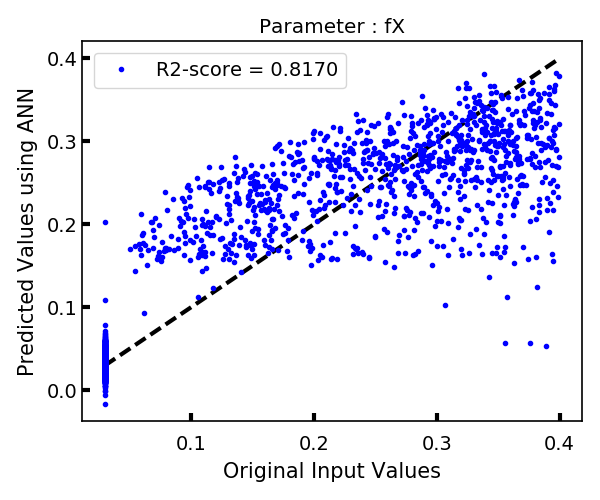}}  
\subfloat[$\rm f_{Xh}$]{\includegraphics[width=0.33\textwidth]{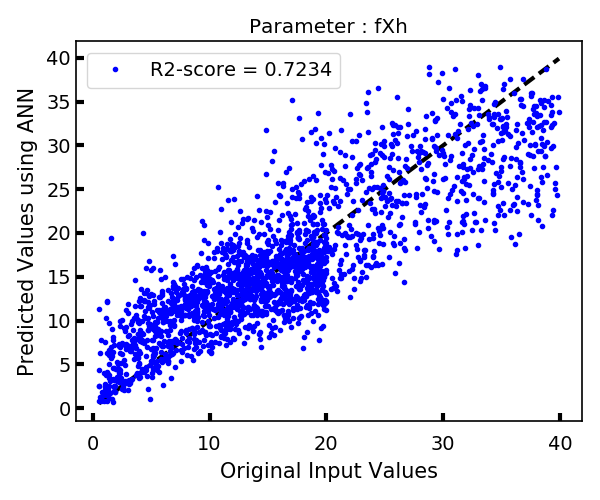}}
\subfloat[$\rm N_{\alpha}$]{\includegraphics[width=0.33\textwidth]{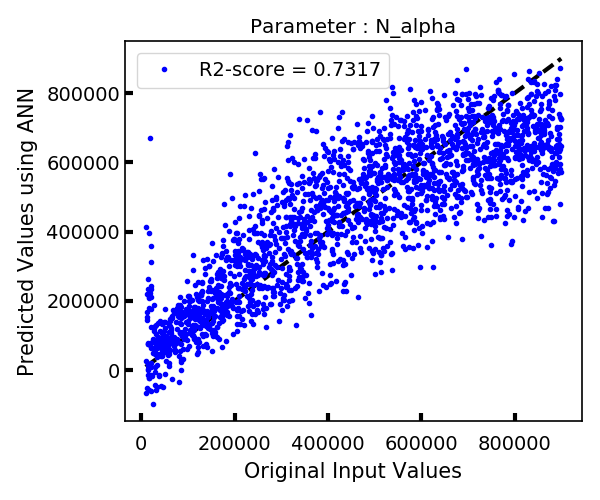}}\\
\subfloat[$\rm f_{star}$]{\includegraphics[width=0.33\textwidth]{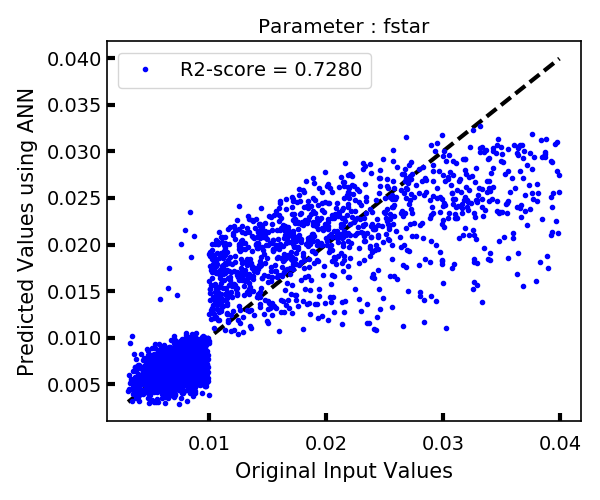}}
\subfloat[$\rm f_R$]{\includegraphics[width=0.33\textwidth]{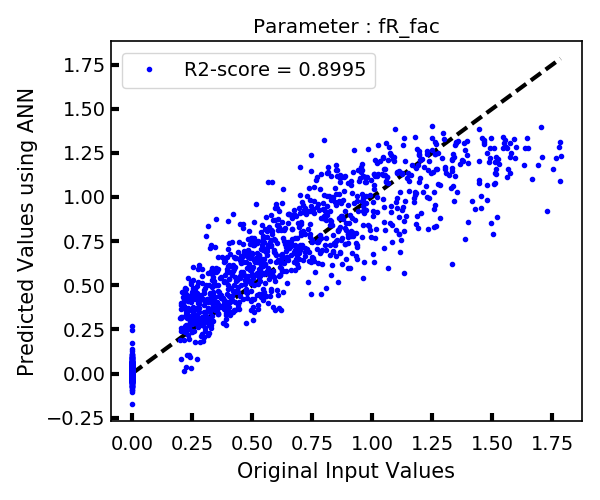}}
\subfloat[$\rm f_{esc}$]{\includegraphics[width=0.33\textwidth]{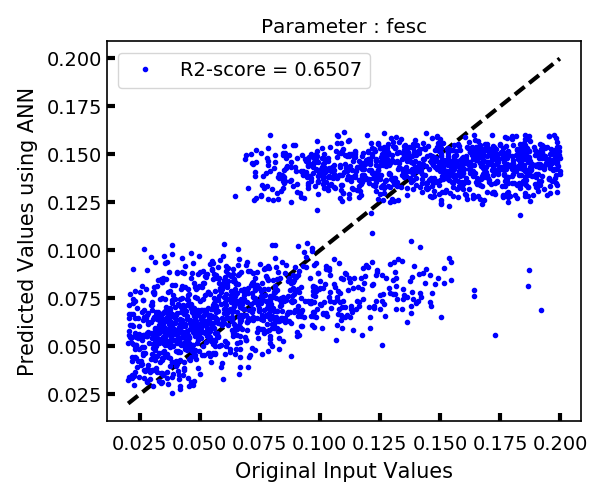}} 

\caption{Each plot above shows the predictions by the network for each parameter, when Test Set A (which contains the 21cm-signals along with foreground and noise) is input. The plots show the original versus predicted values for the signal and foreground parameters. The dashed line through the origin in each plot, represents the true value of the inputs. }
\label{fig:BFG}
\end{figure*}

As described in \textsection\ref{Building_test_sets}, the test set A contains 1000 mock observations(including foreground) unseen to the network. This set is used to evaluate the performance of our network by calculating the $\rm R^2$ scores. We plot the original versus the predicted values of the signal parameters along with the computed $R^2$ score in Fig.~\ref{fig:BFG}. 
We have tabulated these values of $\rm R^2$ for each of the signal parameters in Tab.~\ref{tab:R2scores_all}.
The signal parameters are predicted with $\rm R^2$-scores ranging from 0.65 to 0.81.
\begin{figure}
    \centering
    \includegraphics[width=\columnwidth]{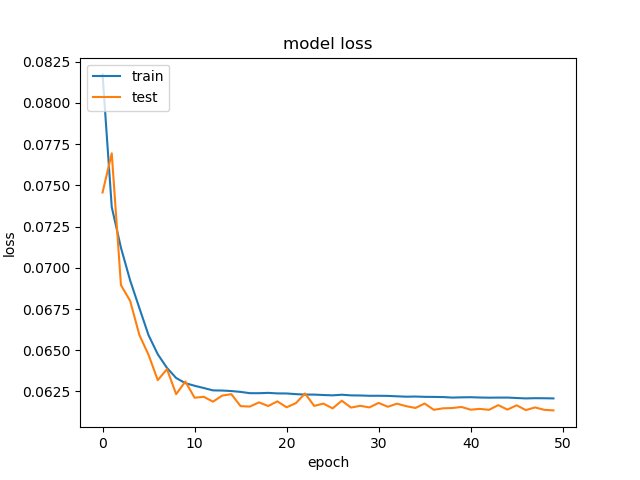}
    \caption{Evolution of the loss function for the network trained by the mock-observations, which includes the signal and the foreground. }
    \label{fig:my_label}
\end{figure}

% \begin{figure*}
% \includegraphics[width=7in, height=5in]{BFG/BFG_predictions.png}
% \caption{The original versus predicted values of the foreground parameters. We observe that the first parameter b0 has the highest R2-score}
%\end{figure*}

\begin{table}
    \centering
    \begin{tabular}{c|c|c}
     \hline
     Parameters   & $\rm R^2$-score(BR-FG)  \\
     \hline
     $\rm f_{*}$   & 0.728\\
     $\rm f_{esc}$   &0.650 \\
     $\rm f_{Xh}$   &0.723\\
     $\rm f_{X}$   &0.816\\
     $\rm f_R$  &0.899\\
     $\rm N_{\alpha}$  & 0.731\\
    \hline
\end{tabular}
\caption{Listing the $\rm R^2$-scores for the all the signal parameters.}
\label{tab:R2scores_all}
\end{table}
\begin{figure*}
\subfloat{\includegraphics[width=\columnwidth]{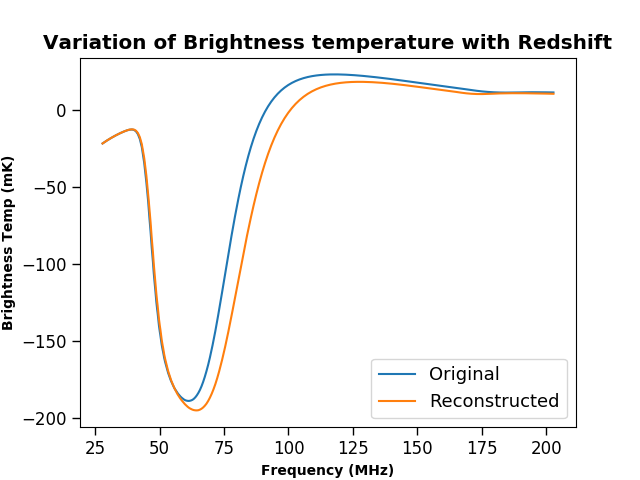}}
\subfloat{\includegraphics[width=\columnwidth]{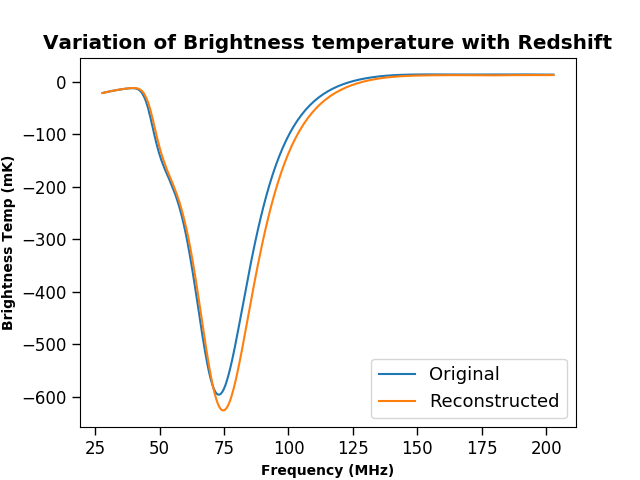}}\\
\subfloat{
\begin{tabular}{c c c c c c}
    \hline
     $\rm f_X$ & $\rm f_{Xh}$ &	$\rm f_R / 10^4$ & $\rm f_*$ &	$\rm f_{esc}$ & $\rm N_{\alpha}$ \\
    \hline
    38.091 &	3e-02 & 0.0 & 	1.756e-02 &	6.948e-02 &	1.6263e+05 \\
    32.484 & 1.29e-02 &	0.0 &	2.141e-02	& 6.754e-02	& 1.1419e+05 \\
    \hline
\end{tabular}}
\subfloat{
\begin{tabular}{c c c c c c}
    \hline
     $\rm f_X$ & $\rm f_{Xh}$ &	$\rm f_R / 10^4$ & $\rm f_*$ & $\rm f_{esc}$ & $\rm N_{\alpha}$ \\
    \hline
    19.0506 & 0.372 &	1.5932 & 5.158e-03	& 0.177 & 4.2925e+05 \\
    13.5237 & 0.310 & 1.1755 & 	5.788e-03 &	0.160 & 2.5537e+05 \\
    \hline
\end{tabular}}
\caption{Reconstructed 21-cm Global signals from the predictions of the ANN when Test set B (which includes two 21-cm signal along with foreground and noise) is input to the ANN. The plot on the left shows the original and the reconstructed signals for the traditional signal input. The plot on the right depicts the case where the input signal is the one with the excess radio background. The original (top row) and predicted (bottom row) values of the parameters are listed in the tables below the plots, respectively. These predicted parameters are used to reconstruct the signal using our model.}
\label{fig:reconstructed}
\end{figure*}

Test set B, consisting of the two known sample signals, (one from the traditional set and the other from the exotic set, along with foregrounds) is then input into the saved network. We use the predicted values of the signal parameters by the network to reconstruct the 21-cm signals using our code. We have shown the original and predicted parameters corresponding to both the samples in the test set B, in a table included below the Fig.~\ref{fig:reconstructed}, along with the plots of the original and reconstructed signals. We can see that the reconstructed signal is similar in shape, to the original signals. The parameter $f_R$ is zero for the traditional signal input and $\sim~11000$ for the model with excess radio background. Also,it is to be noted that the reconstructed signals span the entire frequency range 25-200 MHz, which we can generate using our code, with the input astrophysical parameters. \\
\begin{figure*}
    \centering
    \subfloat{\includegraphics[width=4in,height=3in]{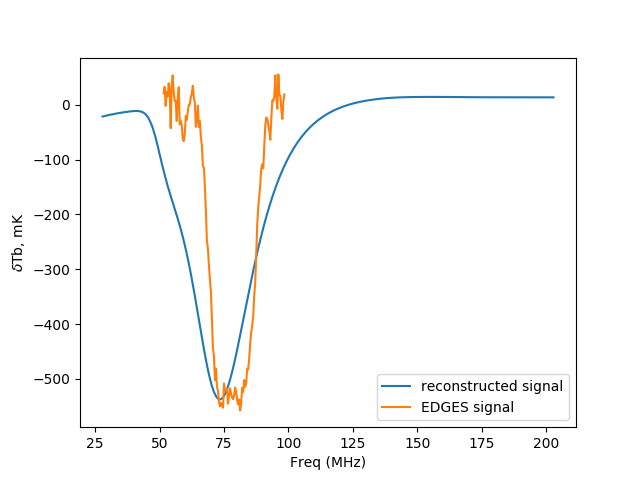}}\\
    \subfloat{
    \begin{tabular}{c c c c c c c}
    \hline
    Parameters & $\rm f_{*}$ & $\rm f_{esc}$ & $\rm f_{Xh}$ & $\rm f_{X}$ & $\rm f_R$ & $\rm N_{\alpha}$ \\
    \hline
    Predicted values & 0.006 & 0.139 & 0.3 & 16.24 & 1e4 & 9728  \\
    \hline
\end{tabular}}
%\caption{Predicted values of the signal parameters, when EDGES data is input to the network.}
%\label{tab:EDGES_predictions}}
   \caption{The reconstructed signal along with the EDGES signal. We can see that by training our ANN with models, which include EDGES-like large absorption troughs apart from the traditional models, we get the predicted parameters which are used to reconstruct the signal. We can see that the magnitude of the signal is comparable to the EDGES detection, but the flattened nature of the EDGES signal cannot be reconstructed by the signal reconstructed by our models.}
\label{fig:EDGESandreconstructed}
\end{figure*}

\subsection{EDGES data as the test input}
The possible detection of the absorption trough in the sky-averaged signal by EDGES, is our only available real observations, currently.  We input the observed sky signal from the EDGES data \footnote{http://loco.lab.asu.edu/edges/edges-data-release/} into the saved network. We pre-process and input the data corresponding to the sub-band 52.7-98.4 MHz to the saved ANN. The network predicts the values of the signal parameters, which is used to reconstruct the signal using our model. We plot the reconstructed signal along with the EDGES signal in Fig.~\ref{fig:EDGESandreconstructed}. The signal parameters predicted by the network are tabulated below the same figure. We see that the amplitude of the signal almost matches with the EDGES signal. The predicted value of the $\rm f_R$ parameter, implies that there is a strong excess radio background, which is responsible for the deep absorption trough of the signal. 

We find that although the reconstructed signal correctly predicts the location and depth of the EDGES signal, the detailed match between the two is not good. This is not surprising as the models where the star formation follows the collapsed fraction of the dark matter halos are unable to account for the flat-bottomed part of the absorption profile and also the rapid rise in the signal in the epoch of heating \citep{EwallWice_2018}. Obtaining a better match would require somewhat non-trivial evolution of the star formation rate, e.g., using PopIII stars \citep{EwallWice_2018, 2020arXiv200305911C}.

\section{Summary and Discussions}
In this work, we have used ANN to estimate the 21-cm signal parameters from synthetic datasets mimicking mock-observations for the first time. Our network is capable of predicting the parameters with $R^2$-scores ranging between 0.65 (for $\rm f_{esc}$) and 0.89 (for $\rm f_R$), when the signal was corrupted by the foregrounds and thermal noise. When we consider the case where there are no foregrounds, the $\rm R^2$score for the combined-parameters ranges between $0.88$ (for $\rm f_X \cdot f_{X,h}$) and 0.96 (for $\rm f_*\cdot f_{esc}$). The corresponding plots showing the predicted and the true values of the parameters are shown in Fig.~\ref{fig:BFG}(for the case where foregrounds are added) and in Fig.~\ref{fig:NoFg}-\ref{fig:NoFg_combined}(where there are no foregrounds). As we are using a single neural network to predict multiple parameters, it is difficult to compare the efficiency of prediction of each parameter individually. The overall accuracy of the neural network deteriorates when the training set is modified by adding foregrounds.
As the 21-cm signal parameters are closely associated with the physics of the evolution of the signal, the prediction of the signal parameters gives us a direct physical interpretation of the signal. The mock observations input to the ANN as test sets, gives us a measure of the reliability of the network. 

The reported EDGES(low-band)-detection with the exceedingly deep absorption trough, could give us a totally new insight on the evolution of the Universe. The parameter $\rm f_R$ determines the magnitude of the radio-excess as described in our model in \textsection~\ref{excess}. From the predictions from our ANN for the simulated mock data sets (which included one traditional signal and one signal with the excess radio background), we observed that the value of $\rm f_R$ contributed most significantly in determining the shape of reconstructed signal. A high value of $\rm f_R$ implies a very bright radio background, resulting in a signal with a large absorption trough, while a $\rm f_R=0$ turns off the excess radio background, and generates a traditional signal. Our network is able to predict the other astrophysical parameters with $\rm R^2$-scores ranging between $\rm 0.65-0.816$, when the input consists of the 21-cm signal along with foregrounds and noise. The parameters $\rm f_{esc}$ and $\rm f_*$ have high negative correlation, possibly explaining their lower $\rm R^2$-scores. The parameters $\rm f_X$ and $\rm f_{X,h}$ determines the location of the minima of the absorption trough, where X-ray heating starts dominating. When we input the EDGES data as the test data, we obtain the  signal parameters as the predictions of the network, which is further used to reconstruct the signal (Fig.~\ref{fig:EDGESandreconstructed}).

The use of ANN to extract the signal parameters efficiently from the total observed sky signal directly without modelling and subtracting the foregrounds, is a novel technique. However, our neural network is limited to the parameter space that we have chosen, and the model foregrounds used. We have used a very specific model of foreground as prescribed by \citet{Bowman_2018}, which is a physically motivated model including the effects of ionosphere. We plan to expand the flexibility of the network by including different models of foreground and introducing  instrumental effects in a future work. These would make the mock-observations more realistic. We observe how crucial the foreground modelling is, in order to retrieve the signal parameters efficiently. Artificial neural networks can very efficiently retrieve the signal parameters when the foreground is removed. 

In contrast to the existing techniques of parameter estimation, ANN can extract features from data by constructing functions which associate the input with the output parameters. They do not require a specified prior, though the training sets can be considered to play a similar role as the prior in Bayesian techniques. The use of ML expedites the computational process considerably. The training process for the  networks we have used for 6000 samples have taken around 12 mins, when trained on a personal computer. Publicly available codes, for example, ARES(Accelerated Reionization Era, \citep{Mirocha_2012,Mirocha_2014} and 21cmFAST \citep{Mesinger_2011} are also  widely used to generate the 21cm Global Signals and reionization simulations. In our code, we have simply introduced an excess-radio background and have not accounted for other exotic phenomena, to generate EDGES-like signals. In this way, we could use a common set of parameters for both the traditional and EDGES-like sets of signals, which enabled us to use a single neural network to extract signal parameters from both types of input signals, even when foregrounds and thermal noise were added. Our code is very simple and is sensitive to each of the astrophysical parameters which it takes as input. We would include a variety of more sophisticated physical models, as incorporated in ARES and 21cmFAST to generate various reionization simulations in future works.  We would like to design a complete signal extraction tool which would be trained with several different models of the signal, foreground and instrument response, such that it could be used as a robust tool to predict the various astrophysical parameters associated with the signal.

\section{Acknowledgements}
MC acknowledges the support of DST for providing the INSPIRE fellowship (IF160153).AD acknowledges the support of EMR-II under CSIR. AC \& TRC acknowledge support of the Department of Atomic Energy, Government of India, under project no. 12-R\&D-TFR-5.02-0700.

\section{Data Availability }
We have used the publicly available EDGES data release, which can be found at \url{http://loco.lab.asu.edu/edges/edges-data-release/}. 
%%%%%%%%%%%%%%%%%%%%%%%%%%%%%%%%%%%%%%%%%%%%%%%%%%



\bibliographystyle{mnras}
\bibliography{References.bib}

\begin{thebibliography}{}
\makeatletter
\relax
\def\mn@urlcharsother{\let\do\@makeother \do\$\do\&\do\#\do\^\do\_\do\%\do\~}
\def\mn@doi{\begingroup\mn@urlcharsother \@ifnextchar [ {\mn@doi@}
  {\mn@doi@[]}}
\def\mn@doi@[#1]#2{\def\@tempa{#1}\ifx\@tempa\@empty \href
  {http://dx.doi.org/#2} {doi:#2}\else \href {http://dx.doi.org/#2} {#1}\fi
  \endgroup}
\def\mn@eprint#1#2{\mn@eprint@#1:#2::\@nil}
\def\mn@eprint@arXiv#1{\href {http://arxiv.org/abs/#1} {{\tt arXiv:#1}}}
\def\mn@eprint@dblp#1{\href {http://dblp.uni-trier.de/rec/bibtex/#1.xml}
  {dblp:#1}}
\def\mn@eprint@#1:#2:#3:#4\@nil{\def\@tempa {#1}\def\@tempb {#2}\def\@tempc
  {#3}\ifx \@tempc \@empty \let \@tempc \@tempb \let \@tempb \@tempa \fi \ifx
  \@tempb \@empty \def\@tempb {arXiv}\fi \@ifundefined
  {mn@eprint@\@tempb}{\@tempb:\@tempc}{\expandafter \expandafter \csname
  mn@eprint@\@tempb\endcsname \expandafter{\@tempc}}}

\bibitem[\protect\citeauthoryear{{Barkana}}{{Barkana}}{2018a}]{Barkana_2018}
{Barkana} R.,  2018a, \mn@doi [\nat] {10.1038/nature25791}, \href
  {http://adsabs.harvard.edu/abs/2018Natur.555...71B} {555, 71}

\bibitem[\protect\citeauthoryear{{Barkana}}{{Barkana}}{2018b}]{barkana2018}
{Barkana} R.,  2018b, \mn@doi [\nat] {10.1038/nature25791}, \href
  {http://adsabs.harvard.edu/abs/2018Natur.555...71B} {555, 71}

\bibitem[\protect\citeauthoryear{{Barkana} \& {Loeb}}{{Barkana} \&
  {Loeb}}{2005}]{Barkana_2005A}
{Barkana} R.,  {Loeb} A.,  2005, \mn@doi [\apjl] {10.1086/430599}, \href
  {http://adsabs.harvard.edu/abs/2005ApJ...624L..65B} {624, L65}

\bibitem[\protect\citeauthoryear{{Bernardi}, {McQuinn}  \&
  {Greenhill}}{{Bernardi} et~al.}{2015}]{Bernardi_2015}
{Bernardi} G.,  {McQuinn} M.,   {Greenhill} L.~J.,  2015, \mn@doi [\apj]
  {10.1088/0004-637X/799/1/90}, \href
  {http://adsabs.harvard.edu/abs/2015ApJ...799...90B} {799, 90}

\bibitem[\protect\citeauthoryear{{Bharadwaj} \& {Ali}}{{Bharadwaj} \&
  {Ali}}{2004}]{bharadwaj2004}
{Bharadwaj} S.,  {Ali} S.~S.,  2004, \mn@doi [\mnras]
  {10.1111/j.1365-2966.2004.07907.x}, \href
  {http://adsabs.harvard.edu/abs/2004MNRAS.352..142B} {352, 142}

\bibitem[\protect\citeauthoryear{{Bowman}, {Rogers}, {Monsalve}, {Mozdzen}  \&
  {Mahesh}}{{Bowman} et~al.}{2018a}]{Bowman_2018}
{Bowman} J.~D.,  {Rogers} A.~E.~E.,  {Monsalve} R.~A.,  {Mozdzen} T.~J.,
  {Mahesh} N.,  2018a, \mn@doi [\nat] {10.1038/nature25792}, \href
  {http://adsabs.harvard.edu/abs/2018Natur.555...67B} {555, 67}

\bibitem[\protect\citeauthoryear{{Bowman}, {Rogers}, {Monsalve}, {Mozdzen}  \&
  {Mahesh}}{{Bowman} et~al.}{2018b}]{bowman2018}
{Bowman} J.~D.,  {Rogers} A.~E.~E.,  {Monsalve} R.~A.,  {Mozdzen} T.~J.,
  {Mahesh} N.,  2018b, \mn@doi [\nat] {10.1038/nature25792}, \href
  {http://adsabs.harvard.edu/abs/2018Natur.555...67B} {555, 67}

\bibitem[\protect\citeauthoryear{{Chardin}, {Uhlrich}, {Aubert}, {Deparis},
  {Gillet}, {Ocvirk}  \& {Lewis}}{{Chardin} et~al.}{2019}]{Chardin_2019}
{Chardin} J.,  {Uhlrich} G.,  {Aubert} D.,  {Deparis} N.,  {Gillet} N.,
  {Ocvirk} P.,   {Lewis} J.,  2019, \mn@doi [\mnras] {10.1093/mnras/stz2605},
  \href {https://ui.adsabs.harvard.edu/abs/2019MNRAS.490.1055C} {490, 1055}

\bibitem[\protect\citeauthoryear{{Chatterjee}, {Dayal}, {Choudhury}  \&
  {Hutter}}{{Chatterjee} et~al.}{2019}]{Chatterjee_2019}
{Chatterjee} A.,  {Dayal} P.,  {Choudhury} T.~R.,   {Hutter} A.,  2019, \mn@doi
  [\mnras] {10.1093/mnras/stz1444}, \href
  {https://ui.adsabs.harvard.edu/abs/2019MNRAS.487.3560C} {487, 3560}

\bibitem[\protect\citeauthoryear{{Chatterjee}, {Dayal}, {Choudhury}  \&
  {Schneider}}{{Chatterjee} et~al.}{2020}]{2020arXiv200305911C}
{Chatterjee} A.,  {Dayal} P.,  {Choudhury} T.~R.,   {Schneider} R.,  2020,
  arXiv e-prints, \href {https://ui.adsabs.harvard.edu/abs/2020arXiv200305911C}
  {p. arXiv:2003.05911}

\bibitem[\protect\citeauthoryear{{Choudhury}, {Datta}  \&
  {Chakraborty}}{{Choudhury} et~al.}{2020}]{Choudhury_2020}
{Choudhury} M.,  {Datta} A.,   {Chakraborty} A.,  2020, \mn@doi [\mnras]
  {10.1093/mnras/stz3107}, \href
  {https://ui.adsabs.harvard.edu/abs/2020MNRAS.491.4031C} {491, 4031}

\bibitem[\protect\citeauthoryear{{Ciardi} \& {Madau}}{{Ciardi} \&
  {Madau}}{2003}]{2003ApJ...596....1C}
{Ciardi} B.,  {Madau} P.,  2003, \mn@doi [\apj] {10.1086/377634}, \href
  {https://ui.adsabs.harvard.edu/\#abs/2003ApJ...596....1C} {596, 1}

\bibitem[\protect\citeauthoryear{{Cohen}, {Fialkov}, {Barkana}  \&
  {Monsalve}}{{Cohen} et~al.}{2019}]{Cohen_2019}
{Cohen} A.,  {Fialkov} A.,  {Barkana} R.,   {Monsalve} R.,  2019, arXiv
  e-prints, \href {https://ui.adsabs.harvard.edu/abs/2019arXiv191006274C} {p.
  arXiv:1910.06274}

\bibitem[\protect\citeauthoryear{Duchi, Hazan  \& Singer}{Duchi
  et~al.}{2011}]{Duchi_2011}
Duchi J.,  Hazan E.,   Singer Y.,  2011, J. Mach. Learn. Res., 12, 2121–2159

\bibitem[\protect\citeauthoryear{{Ewall-Wice}, {Chang}, {Lazio}, {Dor{\'e}},
  {Seiffert}  \& {Monsalve}}{{Ewall-Wice} et~al.}{2018a}]{EwallWice_2018}
{Ewall-Wice} A.,  {Chang} T.-C.,  {Lazio} J.,  {Dor{\'e}} O.,  {Seiffert} M.,
  {Monsalve} R.~A.,  2018a, \mn@doi [\apj] {10.3847/1538-4357/aae51d}, \href
  {http://adsabs.harvard.edu/abs/2018ApJ...868...63E} {868, 63}

\bibitem[\protect\citeauthoryear{{Ewall-Wice}, {Chang}, {Lazio}, {Dor{\'e}},
  {Seiffert}  \& {Monsalve}}{{Ewall-Wice} et~al.}{2018b}]{ewall_wice2018}
{Ewall-Wice} A.,  {Chang} T.-C.,  {Lazio} J.,  {Dor{\'e}} O.,  {Seiffert} M.,
  {Monsalve} R.~A.,  2018b, \mn@doi [\apj] {10.3847/1538-4357/aae51d}, \href
  {http://adsabs.harvard.edu/abs/2018ApJ...868...63E} {868, 63}

\bibitem[\protect\citeauthoryear{{Feng} \& {Holder}}{{Feng} \&
  {Holder}}{2018}]{feng2018}
{Feng} C.,  {Holder} G.,  2018, \mn@doi [\apjl] {10.3847/2041-8213/aac0fe},
  \href {http://adsabs.harvard.edu/abs/2018ApJ...858L..17F} {858, L17}

\bibitem[\protect\citeauthoryear{{Fialkov} \& {Barkana}}{{Fialkov} \&
  {Barkana}}{2019}]{fialkov2019}
{Fialkov} A.,  {Barkana} R.,  2019, arXiv e-prints, \href
  {http://adsabs.harvard.edu/abs/2019arXiv190202438F} {}

\bibitem[\protect\citeauthoryear{Fialkov, Barkana  \& Cohen}{Fialkov
  et~al.}{2018}]{Fialkov_2018}
Fialkov A.,  Barkana R.,   Cohen A.,  2018, \mn@doi [Phys. Rev. Lett.]
  {10.1103/PhysRevLett.121.011101}, 121, 011101

\bibitem[\protect\citeauthoryear{{Field}}{{Field}}{1959}]{Field_1959}
{Field} G.~B.,  1959, \mn@doi [\apj] {10.1086/146653}, \href
  {http://adsabs.harvard.edu/abs/1959ApJ...129..536F} {129, 536}

\bibitem[\protect\citeauthoryear{{Fixsen} et~al.,}{{Fixsen}
  et~al.}{2011}]{fixsen2011}
{Fixsen} D.~J.,  et~al., 2011, \mn@doi [\apj] {10.1088/0004-637X/734/1/5},
  \href {http://adsabs.harvard.edu/abs/2011ApJ...734....5F} {734, 5}

\bibitem[\protect\citeauthoryear{{Fraser} et~al.,}{{Fraser}
  et~al.}{2018}]{fraser2018}
{Fraser} S.,  et~al., 2018, \mn@doi [Physics Letters B]
  {10.1016/j.physletb.2018.08.035}, \href
  {http://adsabs.harvard.edu/abs/2018PhLB..785..159F} {785, 159}

\bibitem[\protect\citeauthoryear{{Furlanetto}, {Oh}  \&
  {Pierpaoli}}{{Furlanetto} et~al.}{2006a}]{furlanetto2006b}
{Furlanetto} S.~R.,  {Oh} S.~P.,   {Pierpaoli} E.,  2006a, \mn@doi [\prd]
  {10.1103/PhysRevD.74.103502}, \href
  {http://adsabs.harvard.edu/abs/2006PhRvD..74j3502F} {74, 103502}

\bibitem[\protect\citeauthoryear{{Furlanetto}, {Oh}  \& {Briggs}}{{Furlanetto}
  et~al.}{2006b}]{Furlanetto_2006}
{Furlanetto} S.~R.,  {Oh} S.~P.,   {Briggs} F.~H.,  2006b, \mn@doi [\physrep]
  {10.1016/j.physrep.2006.08.002}, \href
  {http://adsabs.harvard.edu/abs/2006PhR...433..181F} {433, 181}

\bibitem[\protect\citeauthoryear{{Furlanetto}, {Oh}  \& {Briggs}}{{Furlanetto}
  et~al.}{2006c}]{furlanetto2006c}
{Furlanetto} S.~R.,  {Oh} S.~P.,   {Briggs} F.~H.,  2006c, \physrep, \href
  {http://adsabs.harvard.edu/abs/2006PhR...433..181F} {433, 181}

\bibitem[\protect\citeauthoryear{{Gillet}, {Mesinger}, {Greig}, {Liu}  \&
  {Ucci}}{{Gillet} et~al.}{2019}]{Gillet_2019}
{Gillet} N.,  {Mesinger} A.,  {Greig} B.,  {Liu} A.,   {Ucci} G.,  2019,
  \mn@doi [\mnras] {10.1093/mnras/stz010}, \href
  {https://ui.adsabs.harvard.edu/abs/2019MNRAS.484..282G} {484, 282}

\bibitem[\protect\citeauthoryear{{Greenhill} \& {Bernardi}}{{Greenhill} \&
  {Bernardi}}{2012}]{Greenhill_2012}
{Greenhill} L.~J.,  {Bernardi} G.,  2012, arXiv e-prints, \href
  {https://ui.adsabs.harvard.edu/abs/2012arXiv1201.1700G} {p. arXiv:1201.1700}

\bibitem[\protect\citeauthoryear{{G{\"u}rkan} et~al.,}{{G{\"u}rkan}
  et~al.}{2018}]{gurkan2018}
{G{\"u}rkan} G.,  et~al., 2018, \mn@doi [\mnras] {10.1093/mnras/sty016}, \href
  {http://adsabs.harvard.edu/abs/2018MNRAS.475.3010G} {475, 3010}

\bibitem[\protect\citeauthoryear{{Haardt} \& {Madau}}{{Haardt} \&
  {Madau}}{2012}]{haardt2012}
{Haardt} F.,  {Madau} P.,  2012, \mn@doi [\apj] {10.1088/0004-637X/746/2/125},
  \href {http://adsabs.harvard.edu/abs/2012ApJ...746..125H} {746, 125}

\bibitem[\protect\citeauthoryear{{Hassan}, {Liu}, {Kohn}  \& {La
  Plante}}{{Hassan} et~al.}{2019}]{Hassan_2019}
{Hassan} S.,  {Liu} A.,  {Kohn} S.,   {La Plante} P.,  2019, \mn@doi [\mnras]
  {10.1093/mnras/sty3282}, \href
  {https://ui.adsabs.harvard.edu/abs/2019MNRAS.483.2524H} {483, 2524}

\bibitem[\protect\citeauthoryear{{Hills}, {Kulkarni}, {Meerburg}  \&
  {Puchwein}}{{Hills} et~al.}{2018}]{Hills_2018}
{Hills} R.,  {Kulkarni} G.,  {Meerburg} P.~D.,   {Puchwein} E.,  2018, \mn@doi
  [\nat] {10.1038/s41586-018-0796-5}, \href
  {http://adsabs.harvard.edu/abs/2018Natur.564E..32H} {564, E32}

\bibitem[\protect\citeauthoryear{{Jana}, {Nath}  \& {Biermann}}{{Jana}
  et~al.}{2019}]{jana2019}
{Jana} R.,  {Nath} B.~B.,   {Biermann} P.~L.,  2019, \mn@doi [\mnras]
  {10.1093/mnras/sty3426}, \href
  {http://adsabs.harvard.edu/abs/2019MNRAS.483.5329J} {483, 5329}

\bibitem[\protect\citeauthoryear{{Jennings}, {Watkinson}, {Abdalla}  \&
  {McEwen}}{{Jennings} et~al.}{2019}]{Jennings_2019}
{Jennings} W.~D.,  {Watkinson} C.~A.,  {Abdalla} F.~B.,   {McEwen} J.~D.,
  2019, \mn@doi [\mnras] {10.1093/mnras/sty3168}, \href
  {https://ui.adsabs.harvard.edu/abs/2019MNRAS.483.2907J} {483, 2907}

\bibitem[\protect\citeauthoryear{{Kingma} \& {Ba}}{{Kingma} \&
  {Ba}}{2014}]{Kingma_2014}
{Kingma} D.~P.,  {Ba} J.,  2014, preprint, \href
  {http://adsabs.harvard.edu/abs/2014arXiv1412.6980K} {} (\mn@eprint {arXiv}
  {1412.6980})

\bibitem[\protect\citeauthoryear{{Mebane}, {Mirocha}  \& {Furlanetto}}{{Mebane}
  et~al.}{2020}]{2020MNRAS.493.1217M}
{Mebane} R.~H.,  {Mirocha} J.,   {Furlanetto} S.~R.,  2020, \mn@doi [\mnras]
  {10.1093/mnras/staa280}, \href
  {https://ui.adsabs.harvard.edu/abs/2020MNRAS.493.1217M} {493, 1217}

\bibitem[\protect\citeauthoryear{{Mesinger}, {Furlanetto}  \& {Cen}}{{Mesinger}
  et~al.}{2011}]{Mesinger_2011}
{Mesinger} A.,  {Furlanetto} S.,   {Cen} R.,  2011, \mn@doi [\mnras]
  {10.1111/j.1365-2966.2010.17731.x}, \href
  {http://adsabs.harvard.edu/abs/2011MNRAS.411..955M} {411, 955}

\bibitem[\protect\citeauthoryear{{Mirocha}}{{Mirocha}}{2014}]{Mirocha_2014}
{Mirocha} J.,  2014, \mn@doi [\mnras] {10.1093/mnras/stu1193}, \href
  {http://adsabs.harvard.edu/abs/2014MNRAS.443.1211M} {443, 1211}

\bibitem[\protect\citeauthoryear{{Mirocha} \& {Furlanetto}}{{Mirocha} \&
  {Furlanetto}}{2019}]{mirocha2019}
{Mirocha} J.,  {Furlanetto} S.~R.,  2019, \mn@doi [\mnras]
  {10.1093/mnras/sty3260}, \href
  {http://adsabs.harvard.edu/abs/2019MNRAS.483.1980M} {483, 1980}

\bibitem[\protect\citeauthoryear{{Mirocha}, {Skory}, {Burns}  \&
  {Wise}}{{Mirocha} et~al.}{2012}]{Mirocha_2012}
{Mirocha} J.,  {Skory} S.,  {Burns} J.~O.,   {Wise} J.~H.,  2012, \mn@doi
  [\apj] {10.1088/0004-637X/756/1/94}, \href
  {http://adsabs.harvard.edu/abs/2012ApJ...756...94M} {756, 94}

\bibitem[\protect\citeauthoryear{{Morales} \& {Wyithe}}{{Morales} \&
  {Wyithe}}{2010}]{Morales_2010}
{Morales} M.~F.,  {Wyithe} J.~S.~B.,  2010, \mn@doi [\araa]
  {10.1146/annurev-astro-081309-130936}, \href
  {http://adsabs.harvard.edu/abs/2010ARA%26A..48..127M} {48, 127}

\bibitem[\protect\citeauthoryear{{Nhan}, {Bordenave}, {Bradley}, {Burns},
  {Tauscher}, {Rapetti}  \& {Klima}}{{Nhan} et~al.}{2018}]{Nahn_2018}
{Nhan} B.~D.,  {Bordenave} D.~D.,  {Bradley} R.~F.,  {Burns} J.~O.,  {Tauscher}
  K.,  {Rapetti} D.,   {Klima} P.~J.,  2018, arXiv e-prints, \href
  {https://ui.adsabs.harvard.edu/abs/2018arXiv181104917N} {p. arXiv:1811.04917}

\bibitem[\protect\citeauthoryear{{Patra}, {Subrahmanyan}, {Raghunathan}  \&
  {Udaya Shankar}}{{Patra} et~al.}{2013}]{Patra_2013}
{Patra} N.,  {Subrahmanyan} R.,  {Raghunathan} A.,   {Udaya Shankar} N.,  2013,
  \mn@doi [Experimental Astronomy] {10.1007/s10686-013-9336-3}, \href
  {http://adsabs.harvard.edu/abs/2013ExA....36..319P} {36, 319}

\bibitem[\protect\citeauthoryear{{Pawlik}, {Schaye}  \& {van
  Scherpenzeel}}{{Pawlik} et~al.}{2009}]{pawlik2009}
{Pawlik} A.~H.,  {Schaye} J.,   {van Scherpenzeel} E.,  2009, \mn@doi [\mnras]
  {10.1111/j.1365-2966.2009.14486.x}, \href
  {http://adsabs.harvard.edu/abs/2009MNRAS.394.1812P} {394, 1812}

\bibitem[\protect\citeauthoryear{Pedregosa et~al.,}{Pedregosa
  et~al.}{2011}]{Pedregosa_2011}
Pedregosa F.,  et~al., 2011, Journal of Machine Learning Research, 12, 2825

\bibitem[\protect\citeauthoryear{{Pospelov}, {Pradler}, {Ruderman}  \&
  {Urbano}}{{Pospelov} et~al.}{2018}]{pospelov2018}
{Pospelov} M.,  {Pradler} J.,  {Ruderman} J.~T.,   {Urbano} A.,  2018, \mn@doi
  [Physical Review Letters] {10.1103/PhysRevLett.121.031103}, \href
  {http://adsabs.harvard.edu/abs/2018PhRvL.121c1103P} {121, 031103}

\bibitem[\protect\citeauthoryear{{Pritchard} \& {Loeb}}{{Pritchard} \&
  {Loeb}}{2010}]{Pritchard_2010}
{Pritchard} J.~R.,  {Loeb} A.,  2010, \mn@doi [\prd]
  {10.1103/PhysRevD.82.023006}, \href
  {http://adsabs.harvard.edu/abs/2010PhRvD..82b3006P} {82, 023006}

\bibitem[\protect\citeauthoryear{{Pritchard} \& {Loeb}}{{Pritchard} \&
  {Loeb}}{2012a}]{Pritchard_2012}
{Pritchard} J.~R.,  {Loeb} A.,  2012a, \mn@doi [Reports on Progress in Physics]
  {10.1088/0034-4885/75/8/086901}, \href
  {http://adsabs.harvard.edu/abs/2012RPPh...75h6901P} {75, 086901}

\bibitem[\protect\citeauthoryear{{Pritchard} \& {Loeb}}{{Pritchard} \&
  {Loeb}}{2012b}]{pritchard2012}
{Pritchard} J.~R.,  {Loeb} A.,  2012b, \mn@doi [Reports on Progress in Physics]
  {10.1088/0034-4885/75/8/086901}, \href
  {http://adsabs.harvard.edu/abs/2012RPPh...75h6901P} {75, 086901}

\bibitem[\protect\citeauthoryear{{Pritchard} et~al.,}{{Pritchard}
  et~al.}{2015}]{Pritchard_2015}
{Pritchard} J.,  et~al., 2015, Advancing Astrophysics with the Square Kilometre
  Array (AASKA14), \href {http://adsabs.harvard.edu/abs/2015aska.confE..12P}
  {p.~12}

\bibitem[\protect\citeauthoryear{{Schmit} \& {Pritchard}}{{Schmit} \&
  {Pritchard}}{2018}]{Schmit_2017}
{Schmit} C.~J.,  {Pritchard} J.~R.,  2018, \mn@doi [\mnras]
  {10.1093/mnras/stx3292}, \href
  {http://adsabs.harvard.edu/abs/2018MNRAS.475.1213S} {475, 1213}

\bibitem[\protect\citeauthoryear{{Shimabukuro} \& {Semelin}}{{Shimabukuro} \&
  {Semelin}}{2017}]{Shimabukuro_2017}
{Shimabukuro} H.,  {Semelin} B.,  2017, \mn@doi [\mnras]
  {10.1093/mnras/stx734}, \href
  {http://adsabs.harvard.edu/abs/2017MNRAS.468.3869S} {468, 3869}

\bibitem[\protect\citeauthoryear{{Singh} et~al.,}{{Singh}
  et~al.}{2017}]{Singh_2017}
{Singh} S.,  et~al., 2017, \mn@doi [\apjl] {10.3847/2041-8213/aa831b}, \href
  {http://adsabs.harvard.edu/abs/2017ApJ...845L..12S} {845, L12}

\bibitem[\protect\citeauthoryear{{Slatyer} \& {Wu}}{{Slatyer} \&
  {Wu}}{2018}]{slatyer2018}
{Slatyer} T.~R.,  {Wu} C.-L.,  2018, \mn@doi [\prd]
  {10.1103/PhysRevD.98.023013}, \href
  {http://adsabs.harvard.edu/abs/2018PhRvD..98b3013S} {98, 023013}

\bibitem[\protect\citeauthoryear{{Sokolowski} et~al.,}{{Sokolowski}
  et~al.}{2015}]{Sokolowski_2015}
{Sokolowski} M.,  et~al., 2015, \mn@doi [\pasa] {10.1017/pasa.2015.3}, \href
  {http://adsabs.harvard.edu/abs/2015PASA...32....4S} {32, e004}

\bibitem[\protect\citeauthoryear{{Voytek}, {Natarajan}, {J{\'a}uregui
  Garc{\'{\i}}a}, {Peterson}  \& {L{\'o}pez-Cruz}}{{Voytek}
  et~al.}{2014}]{Voytek_2014}
{Voytek} T.~C.,  {Natarajan} A.,  {J{\'a}uregui Garc{\'{\i}}a} J.~M.,
  {Peterson} J.~B.,   {L{\'o}pez-Cruz} O.,  2014, \mn@doi [\apjl]
  {10.1088/2041-8205/782/1/L9}, \href
  {http://adsabs.harvard.edu/abs/2014ApJ...782L...9V} {782, L9}

\bibitem[\protect\citeauthoryear{{Wouthuysen}}{{Wouthuysen}}{1952}]{Wouthuysen_1952}
{Wouthuysen} S.~A.,  1952, \mn@doi [\aj] {10.1086/106661}, \href
  {http://adsabs.harvard.edu/abs/1952AJ.....57R..31W} {57, 31}

\makeatother
\end{thebibliography}

% if your bibtex file is called example.bib

% \begin{thebibliography}{99}
% \bibitem[\protect\citeauthoryear{Author}{2012}]{Author2012}
% Author A.~N., 2013, Journal of Improbable Astronomy, 1, 1
% \bibitem[\protect\citeauthoryear{Others}{2013}]{Others2013}
% Others S., 2012, Journal of Interesting Stuff, 17, 198
% \end{thebibliography}

%%%%%%%%%%%%%%%%%%%%%%%%%%%%%%%%%%%%%%%%%%%%%%%%%%

%%%%%%%%%%%%%%%%% APPENDICES %%%%%%%%%%%%%%%%%%%%%

% \appendix

% \section{Some extra material}

% If you want to present additional material which would interrupt the flow of the main paper,
% it can be placed in an Appendix which appears after the list of references.

%%%%%%%%%%%%%%%%%%%%%%%%%%%%%%%%%%%%%%%%%%%%%%%%%%

% Don't change these lines
\bsp	% typesetting comment
\label{lastpage}
\end{document}